\documentclass{article}
\title{F and M Theories as Gauge Theories\\
of Area Preserving Algebra}
\author{Hirotaka Sugawara \\
 KEK, High Energy Accelerator Research Organization \\
 Tsukuba, Ibaraki, Japan}

\begin{document}
\maketitle
\begin{abstract}
F theory and M theory are formulated as gauge theories of area preserving 
diffeomorphism algebra. Our M theory is shown to be 1-brane formulation 
rather than 0-brane formulation of M theory of Banks, Fischler, Shenker and 
Susskind and the F theory is shown to be 1-brane 
formulation rather than -1-brane formulation of type IIB matrix theory of 
Ishibashi, Kawai, Kitazawa and Tsuchiya.
\end{abstract}

\paragraph{Area preserving diffeomorphism algebra\\}
Our starting point is the following 2 + 1 dimensional energy-momentum 
algebra of Dirac$^{(1)}$,  Schwinger$^{(2)} $ and DeWitt$^{(3)}$.

\begin{eqnarray}
\left[ T_{00}(\sigma), T_{00} (\sigma') \right] &=& -i\left(T_{i0}(\sigma) + 
T_{i0} (\sigma') \right) \partial_i \delta(\sigma - \sigma') \\
\left[ T_{i0}(\sigma),T_{00}(\sigma') \right] &=& -i T_{00}(\sigma)\partial 
_i \delta(\sigma-\sigma'), \\
\left[ T_{i0}(\sigma), T_{j0}(\sigma') \right] &=& -
i(T_{i0}(\sigma')\partial_j\delta
(\sigma-\sigma')+ T_{j0}(\sigma)\partial_i\delta
(\sigma-\sigma')),
\end{eqnarray}

where $$\sigma = (\sigma^1, \sigma ^2), (i,j=1,2)$$ and 
$$\partial_i\delta(\sigma-\sigma') = \frac{\partial}{\partial\sigma_i}[\delta(\sigma_1 -
\sigma'_1)\delta(\sigma_2 - \sigma'_2)].$$

These commutation relations are valid in an arbitrary Riemannian space 
where the indices 1 and 2 describe a time-like plane and the index 0 
corresponds to the normal direction to the plane.

The area preserving diffeomorphism is defined as
\begin{equation}
T(\sigma)=\frac{1}{\sqrt{g}}(\partial_1T_{20}(\sigma) - \partial_2T_{10}
(\sigma)),
\end{equation}
where $$g = det g_{ij}(\sigma)$$

$g_{ij}(\sigma)$ is the metric tensor of the given time-like plane.

A straightforward calculation leads to the following commutation relation;
\begin{equation}
[T(\sigma), T(\sigma')] = \frac{-
i}{\sqrt{gg'}}\left( \partial_1T(\sigma)\partial
_2\delta(\sigma - \sigma') - \partial_2T(\sigma)\partial_1\delta(\sigma 
- \sigma') \right).
\end{equation}

This commutation relation plays a fundamental role in our formulation of 
F and M theory which are basically membrane theories$^{(4)}$. 

Corresponding to each two dimensional field $A(\sigma)$ we can define the 
following operator $A$;
\begin{equation}
A = \int A(\sigma)T(\sigma)\sqrt{g}d{\sigma_1}d{\sigma_2}
\end{equation}
Then for two operators $A$ and similar operator $B$ we get,
\begin{equation}
[A,B] = 
i\int\{A(\sigma),B(\sigma)\}T(\sigma)\sqrt{g}d\sigma_1d\sigma_2
\end{equation}
with
\begin{equation}
\{A(\sigma),B(\sigma)\}\equiv \frac{1}{\sqrt{g}}
  \left(
  \frac{\partial A}{\partial\sigma_1}\frac{\partial B}{\partial\sigma_2} 
-  \frac{\partial A}{\partial\sigma_2}\frac{\partial B}{\partial\sigma_1}
  \right).
\end{equation}
  
Equation (7) describes the most imortant property of our diffeomorphism 
algebra which relates directly to the gauge theory of the diffeomorphism to the 
theory of membranes.\\
 It is easy to see that $T(\sigma)$ generates the area preserving 
diffeomorphism in the following way. 
 Consider the transformation

\begin{equation}
U[f] = e^{i\int f(\sigma)T(\sigma)\sqrt{g}d\sigma_1d\sigma_2}
\end{equation}
Straightforward calculation gives:\\
if 
\begin{equation}
A' = U^{-1}AU,
\end{equation}
 then 
\begin{equation}
  A'(\sigma) = -\sqrt{g}\{f(\sigma),A(\sigma)\}
\end{equation}
for the infinitesimal function $f(\sigma)$.

The relation to the area preserving diffeomorphism studied previously$^{(5)\sim(10)}$ is as follows.

Let $\{Y_{nm}(\sigma)\}$ be the complete set of functions on a given 
surface S. Then T$(\sigma)$ can be expanded in the following way;

\begin{equation}
T(\sigma) = \sum_{n,m} T_{nm}Y_{nm}(\sigma).
\end{equation}  

With the following normalization and the completeness condition,
\begin{eqnarray}
\int Y_{nm}(\sigma)Y^\ast_{n'm'}(\sigma)\sqrt{g}d\sigma_1d\sigma_2 
&=& \delta_{nn'}\delta_{mm'}, \\
\sum_{n,m}Y_{nm}(\sigma)Y^\ast_{nm}(\sigma') &=& 
\frac{1}{\sqrt{g}}\delta(\sigma - \sigma'),
\end{eqnarray}
we obtain
\begin{equation}
[T_{nm},T_{n'm'}] = 
i\int\{Y^\ast_{nm}(\sigma),Y^\ast_{n'm'}\}T(\sigma)\sqrt{g}d\sigma 
_1d\sigma_2.
\end{equation}

This commutation relation is equivalent to that of the area preserving 
diffeomorphism studied earlier. $Y_{nm}(\sigma)$ is the ordinary spherical 
harmonics $Y_{lm} 
(\theta, \phi)$ in the case of two dimensional 
sphere and the equation $(15)$ reduces to
\begin{equation}
[T_{lm}, T_{l'm'}] = ig_{lm, l'm'}^{l''m''}T_{l"m"},
\end{equation}
where $g_{lm, l'm'}^{l''m''}$ is defined in reference(4). 

As was shown by de Wit, Hoppe and Nicolai$^{(4)}$  this reduces to the 
SU(N) algebra when truncated to $l \leq N-1$. This in turn means that 
$\lim_{N\to\infty}SU(N)$ coincides with our area preserving algebra 
defined through the equation $(5)$ in the special case of 2-dimensional 
sphere.
We also get $w_\infty^{(11)}$ algebra in the case of torus:
\begin{equation}
[T_{\vec{m}}, T_{\vec{m}'}] = \frac{-
i}{4\pi^2}(\vec{m}\times\vec{m}')T_{(\vec{m}+\vec{m}')}
\end{equation}

where $\vec{m}\times\vec{m}' = m_1m'_2-m'_1 m_2$.

\paragraph{Gauge theory of area preserving diffeomorphism 
algebra}$^{(12)\sim(15)}$\\
$X^\mu (\mu=0,\ldots, D-1)$ stands for the D dimensional coordinate and 
$X^\mu (\rho)$ with $\rho = (\rho_1, \ldots, \rho_p)$ describes a  $p$ 
 dimensional space. A gauge field $A_\mu (X(\rho))$ 
 is a function of 
$\rho$ through $X^\mu {(\rho)}$ (not a functional) and is an element of area 
preserving diffeomorphism algebra:\\

\begin{equation}
A_\mu(X(\rho)) = \int A_\mu(X(\rho), \sigma) T(\sigma) \sqrt{g} 
d\sigma_1 d\sigma_2.
\end{equation}

D-dimensional Majorana spinor is also introduced in a similar manner:\\
\begin{equation}
\psi(X(\rho)) = \int\psi(X(\rho), \sigma) T(\sigma) \sqrt{g}d\sigma_1  
d\sigma_2.
\end{equation}

The Lagrangian for the gauge theory of area preserving diffeomorphism  
algebra is given by:

\begin{equation}
L= \frac{-1}{4g^2}[T_rF_{\mu \nu}F^{\mu \nu}+2iT_r \overline{\psi} 
\gamma^{\mu} [D_\mu, \psi ]]
\end{equation}

with 
\begin{equation}
D_\mu = \frac{\delta}{\delta X^\mu(\rho)} + iA_\mu(X(\rho)),
\end{equation}

where the trace will be calculated using\\

\begin{equation}
Tr T(\sigma)T(\sigma')=\frac{1}{\sqrt g}\delta(\sigma-\sigma'),
\end{equation}

and $T_r T(\sigma)=0$.\\
The Lagrangian is invariant under\\

\begin{equation}
A'_\mu=U^{-1}A_\mu U-iU^{-1}\frac{\delta}{\delta X^\mu(\rho)}U,
\end{equation}
and
\begin{equation}
\psi'=U^{-1}\psi U
\end{equation}

where U is defined in the equation $(9)$.\\
The action integral is given by:\\

\begin{equation}
S= \int L d\Omega_\rho,
\end{equation}

where $d \Omega_\rho$  is an appropriate measure in the $\rho$-space.

\paragraph{Example 1. M theory\\}
In this case D=11 with the metric $(- + \cdots+)$. $X^\mu(\rho)$ is chosen to be\\
\begin{equation}
\left\{
    \begin {array}{l}
      X^0=\tau,\\
      X^{10}=\rho,\\
      X^1=X^2=\cdots X^9=0\\
    \end{array} \right.
\end{equation}

This choice of $X^\mu(\rho)$ can be interpreted to imply that the 10th 
space-like coordinate $X^{10}$ is compactified.$^{(16)} $\\
This theory is now shown to coincide with the M-theory formulated by Banks, 
Fischler , Shenker and Susskind,$^{(17)}$ and therefore, with the membrane theory 
of deWit, Hoppe and Nicolai$^{(4)} $in a certain limit.\\
First define$X^\pm=X^{10}\pm X^0 $ and  $A_\pm=\frac{1}{\sqrt{2}}(A_0\pm 
A_{10})$.

The gauge condition $(A_{-} = 0)$ will be imposed from here on.\\
This gauge condition will be supplemented as in the case of Green-Schwarz formulation of the superstring theory by the condition,\\

\begin{equation}
\left(\frac{\delta X^\mu}{\delta\tau} -\frac{\delta  
X^\mu}{\delta\rho}\right)\alpha_{\mu}\psi=0 ,
\end{equation}

where $\alpha_{\mu}$ is defined as usual:\\

\begin{equation}
\alpha_0=\beta=\gamma_0, i\alpha_a = \gamma_0\gamma_a (a=1,\ldots, 10).
\end{equation}

The meaning of this condition will become clear shortly.
To eliminate the $A_+$ component entirely from the Lagrangian we go to 
the light cone system by the following Lorentz transformation;

\begin{equation}
\left(\begin{array}{l}
  X_{0} \\
  X_{10}
\end{array}\right)
=\gamma
\left(\begin{array}{rr}
  1 & -v \\
  -v & 1
\end{array}\right)
\left(\begin{array}{l}
  X_{0}' \\
  X_{10}'
\end{array}\right)
\end{equation}

with $\gamma= \frac{1}{\sqrt{1-\upsilon^2}}$
and take the limit $\upsilon\rightarrow1.$\\
The bosonic part of the Lagrangian then becomes\\

\begin{equation}
L_B=-\frac{T_r}{4g^2}\left[4(\partial_{+} A_a\partial_-A _a)-
(\left[A_a,A_b\right])^2\right],
\end{equation}

with  \(a,b = 1,\cdots, 9\).\\
This is easily shown to be essentially equivalent to the bosonic part of the 
Lagrangian of deWit, Hoppe and Nicolai$^{(4)} $with the help of the equations 
$(7)$ and $(22)$ in the limit@$R_{10} \rightarrow 0$.\\
The vanishing compactification radius \(R_{10}\) can be taken to imply that 
our 1-brane formulation goes to the 0-brane formulation of Banks et al $^{(17)}$.\\
$(\alpha_0, \ldots, \alpha_{10})$ defined in equation $(28)$ generate a 
Clifford algebra just as the original $\gamma_\mu$. The spinorial 
condition $(27)$ can be written as:\\
\begin{equation}
\gamma_{-}\psi = 0,
\end{equation}
where

\begin{equation}
\gamma_{\pm} = \frac{1}{\sqrt{2}}(\beta\pm\alpha_{10}) = \frac{\gamma_0}{
\sqrt{2}}(1\mp 
i\gamma_{10}) = \frac{\gamma_0}{\sqrt{2}}(1\mp\gamma_{ch})
\end{equation}

and where $\gamma_{ch}$ is the 10-dimensional chiral operator. This implies 
that the light-cone condition in the $(\beta, \alpha)$ algebra is the chirality 
condition in the original $\gamma_\mu$ algebra. This condition, therefore, 
guarantees the invariance of our Lagrangian under the following 
supersymmetric transformation:

\begin{equation}
\delta A_\mu=i \overline{\varepsilon}\gamma_\mu\psi,
  \delta\psi_\alpha=\frac{1}{2} 
F_{\mu\nu}(\gamma^{\mu\nu}\epsilon)_\alpha
+ \varepsilon'_\alpha
\end{equation}

The solution to equation $(31)$ can be written as:\\

\begin{equation}
\psi= \psi_+= \frac{1}{{\sqrt{2}i}}\gamma_-\psi_-,    
\end{equation}
@
 with     $\gamma_+\psi_-=0.$

The whole Lagrangian now takes the form\\

\begin{equation}
L = \frac{-T_r }{4g^2}\left[4(\partial_+A_a\partial_-A_a)
                  -([A_a, A_b])^2 + \sqrt{2}i\overline{ \psi_{-}} \left [ \left\{ \gamma_-
                   [D_0,\psi_-] 
                  + \gamma_{-}\alpha^a
                  [A_a,\psi_-] \right\} \right] \right].
\end{equation}

The \(R_{10}\rightarrow 0\) limit of this Lagrangian is nothing but the M-
theory Lagrangian proposed by Banks et al$^{(17)}$, based on the work of 
deWit et al.$^{(4)}$

\renewcommand{\thefootnote}{\fnsymbol{footnote}}
\paragraph{Example 2. F theory$^{(18)}$}\footnote{We are currently working on F-theory based on the volume preserving 
diffeomorphism. Let us call tentatively the 12-dimensional theory in this paper an F-theory simply because it is a 12-dimensional theory which 
leads to the type IIB string theory (to be published).}

The 12 dimensional spinor with the metric $(g_{0,0} ,g_{1,1}, \cdots ,g_{10,10},
g_{11,11})$ $= (-1,+1,
\cdots,+1,-1)$ can have the Majorna-Weyl representation which is 32 dimensional. In 
addition to the 11-dimensional $\gamma$-matrices 
\(\gamma_\mu(\mu=0, \cdots, 10)\) we add $\gamma_{11}=1$ to 
construct the 12-dimensional gauge theory. The Lagrangian is formally the 
same as in the equation $(20)$\\

\begin{equation}
L=-\frac{T_r }{4g^2}(F_{\mu\nu}F^{\mu\nu}+2i\bar\psi \gamma^\mu 
[D_\mu,\psi]),
\end{equation}

where $$\mu,\nu=0,1,\cdots,10,11$$ and $\psi$ is a 32 component 
spinor.\\
$X^\mu(\sigma)$ in this case is defined by\\
$
     X^{11}=\tau,
     X^{10}=\rho,$ and all the other$X^\mu = 0,$\\

and the spinor is assumed to satisfy\\

\begin{equation}
(i\frac{\delta X ^\mu}{\delta \rho}-\frac{\delta X^\mu}{\delta 
\tau})\gamma_\mu\psi=0.
\end{equation}

We define
\begin{equation}
A_\pm=\frac{1}{\sqrt2}(A_{11}-A_{10})
\end{equation}
and  \(X\pm=X^{10} \pm X^{11}\). We take the $A_{-}=0$ gauge and go to 
the light cone gauge by the transformation defined in equation $(29)$ expect 
that $X^0$ is replaced by $X^{11}$.\\
We then obtain\\

\begin{equation}
L = \frac{-Tr}{4g^2}
\left[
  4g^{\mu\nu}\partial_{+}A_{\mu}\partial_{-}A_\nu - g^{\mu\nu} g^{\rho \kappa}
  [A_\rho, A_\mu][A_\kappa, A_\nu] - 2\bar\psi\gamma^\mu [A_\mu,\psi]
\right],
\end{equation}

 where $\mu,\nu=0,\cdots,9$.\\
This Lagrangian coincides with that of Ishibashi, Kawai, Kitazawa and 
Tsuchiya$^{19}$ when the compactification radii in both 11 and 10 directions are 
taken to vanish.\\

\paragraph{Example 3. \(SU(\infty)\) QCD\\}
In this case\\
\begin{eqnarray}
L = \frac{-1}{4g^2}
\left[
  [T_r F_{\mu\nu}F^{\mu\nu} - 2iT_r \bar\psi \gamma^\mu[D_\mu,\psi]
\right] \nonumber
\end{eqnarray}
with $\mu,\nu=0,1,2,3$ and $X^\mu=\rho^\mu$.  We use the notation $X^\mu$ rather than $\rho^\mu$ in this case.\\
The vacuum equation is
$F_{\mu\nu}=0.$\\

The solution to this equation will satisfy\\
\begin{eqnarray}
\oint_C A_{\mu} dX^{\mu}& =& -i\int[A_\mu, A_\nu] dX^\mu 
dX^\nu\nonumber\\
                     & =&\frac12 \int_S \varepsilon_{\alpha\beta}
                       \varepsilon_{ijk} \int_{\Sigma} \frac{T(\sigma)}
{(n,A)} A_i \partial_\alpha A_j \partial_\beta A_k d\sigma_1 d\sigma_2
dS,
\end{eqnarray}

where $A_0 = 0$ gauge was taken.  $dS$ is the surface element of $S$ which
has the boundary $C$. $\Sigma$ is an internal sphere parametrized by $(\sigma_1, 
\sigma_2)$.    
$(n, A)$ is the component of $A$ in the normal direction to the surface $S$.

Following Arafune et al $^{(20)}$ we can rewrite this in the following way:\\

\begin{equation}
\oint_C A_\mu dX^\mu = d\int_S \int_\Sigma \frac{T(\sigma)}{(n,A)}
\sqrt{G} d\sigma_1 d\sigma_2 dS,
\end{equation}
where $G= det(\frac{\partial Ai}{\partial\sigma_\alpha}\frac{\partial Ai}{\partial\sigma_\beta})$ and $d$  is the wrapping number.\\
In contrast to the case of Arafune at al$^{(20)}$ the internal space and the external 
space are interchanged in our case.\\
The vacuum of $SU(\infty) QCD,$ therefore, is the state where the T'Hooft-
Polyakov$^{(21)(22)} $ monopole in the internal space is condensed.\\

\paragraph{Super area preserving Algebra\\}
As in the case of string theory where we have the light-cone formulation 
(Green-Schwartz formulation) on one hand and the Neveu-Schwartz-Ramond 
formulation with the world sheet supersymmetry on the other, our
formulation of F and M theory should also be supplemented by another 
formulation which keeps the world sheet supersymmetry.
For this purpose the two dimensional surface $(\sigma_1, \sigma_2)$ will 
be extended to the super surface parametrized by $(\sigma_1, \sigma_2, 
\theta_1, \theta_2)$ where $\theta_\alpha (\alpha=1, 2)$  is the two 
component spinor with the Grassmannian property. 
The commutation relation $(5)$  is now extended to:

\begin{eqnarray}
[T(\sigma, \theta), T(\sigma', \theta')] &=& \frac{-i}{\sqrt{gg'}}
\left\{
  \frac{\partial T}{\partial\sigma_1} \partial_2 
    \delta (\sigma - \sigma') \delta (\theta - \theta')
\right. \nonumber \\
- \frac{\partial T}{\partial\sigma_2} \partial_{1}
    \delta (\sigma - \sigma') \delta (\theta - \theta')
&-& 
\left .
\frac{\partial T}{\partial\theta_\alpha}
    \frac{\partial}{\partial\theta_\alpha}
    \delta (\sigma - \sigma') \delta (\theta - \theta')
\right\}
\end{eqnarray}

It is easy to prove that for\\
\begin{eqnarray}
A_{\mu} &=& \int \sqrt{g} d\sigma_1 d\sigma_2 d\theta_1 d\theta_2 
A_\mu (\sigma, \theta) T(\sigma, \theta),\\
\left[A_\mu, A_\nu\right] &=& i\int d\Omega \frac{T(\sigma, 
\theta)}{\sqrt{g}}\left(\frac{\partial A_\mu}{\partial\sigma_1} \frac{\partial 
A_\nu}{\partial\sigma_2}
-\frac{\partial A_\mu}{\partial\sigma_2}\frac{\partial A_\nu}{\partial
\sigma_1}-\frac{\partial A_\mu}{\partial\theta_\alpha}\frac{A_\nu}{\partial\theta_\alpha}\right),
\end{eqnarray}

where $d\Omega = \sqrt{g} d\sigma_1 d\sigma_2 d\theta_1 d\theta_2$.

This is consistent with the graded area preserving algebra studied 
earlier$^{(9)}$.

By expanding the $T(\sigma, \theta)$:
\begin{equation}
T(\sigma, \theta) = S(\sigma) + V_\alpha (\sigma) \theta_\alpha + 
T(\sigma) \theta_1 \theta_2\nonumber\\,
\end{equation}

the commutation relation $(42)$ can be rewritten using these component 
fields,

\begin{eqnarray}
[S(\sigma), S(\sigma')] &=& 0,\\
\left[
  V_\alpha (\sigma), S(\sigma')
\right] &=& \frac{-i}{\sqrt{gg'}} 
\varepsilon_{\alpha \beta} V_{\beta}(\sigma) \delta(\sigma-\sigma'),\\
       \left[T(\sigma), S(\sigma')\right] &=& \frac{-i}{\sqrt{gg'}} 
\left(\frac{\partial S(\sigma)} {\partial \sigma_{1}}\partial_2\delta(\sigma-\sigma')-
\frac{\partial S(\sigma)}{\partial\sigma_2} \partial_1 \delta(\sigma-
\sigma')\right)\\
\left[V_\alpha (\sigma), V_\beta (\sigma')\right] &=& \frac{-i}{\sqrt{gg'}}
\left(\frac{\partial S(\sigma)} {\partial \sigma_{1}}\delta(\sigma-\sigma')-
\frac{\partial S(\sigma)}{\partial\sigma_2} \partial_1 \delta(\sigma-
\sigma')\right) \varepsilon_{\alpha \beta}\nonumber\\
&&{ } + \frac{-i}{\sqrt{gg'}}T(\sigma) \delta(\sigma-\sigma') 
\delta_{\alpha \beta},\\
\left[V_\alpha (\sigma), T(\sigma')\right] &=& \frac{-i}{\sqrt{gg'}}
\left(\frac{\partial V_\alpha(\sigma)} {\partial\sigma_{1}} \partial_{2}\delta(\sigma-
\sigma')-\frac{\partial V_\alpha (\sigma)}{\partial\sigma_2} 
\partial_1 \delta(\sigma-\sigma')\right),
\end{eqnarray}
and \\
\begin{eqnarray}
\left[T(\sigma), T(\sigma')\right] &=& \frac{-i}{\sqrt{gg'}}
\left(\frac{\partial T(\sigma)}{\partial \sigma_1}\partial_2 \delta 
(\sigma-\sigma') -\frac{\partial T(\sigma)}{\partial\sigma_2} \partial_1 
\delta(\sigma-\sigma')\right).
\end{eqnarray}

$T(\sigma)$ coincides with the original $T(\sigma)$, and $V_\alpha$ is 
expected to be written in terms of the $2+1$ supercurrent. The gauge theory 
based on the graded algebra $(42)$ is expected to preserve the space-time 
supersymmetry after the GSO projection. 

The discussions with K. Higashijima and N. Ishibashi are greatly 
appreciated.\\

\end{document}